%% file: SelfConsistentAxino.tex
\documentclass[floatfix,twocolumn,superscriptaddress,a4paper,prl,preprintnumbers]{revtex4}

\usepackage{amsmath}
\usepackage{amssymb}
\usepackage{graphicx}
\usepackage{ulem}
\usepackage{color}
\usepackage[T1]{fontenc} 
\usepackage{slashed}
\usepackage{xspace}
\usepackage{hyperref}

\newcommand{\AddrBonn}{%
Bethe Center for Theoretical Physics \& Physikalisches Institut der 
Universit\"at Bonn, \\
 53115 Bonn, Germany
}

\preprint{BONN-TH-2014-04}
\begin{document}

\title{From the unification scale to the weak scale: A self consistent
  supersymmetric Dine-Fischler-Srednicki-Zhitnitsky axion model}

 \author{Herbi K. Dreiner} \email{dreiner@th.physik.uni-bonn.de}
 \affiliation{\AddrBonn}

 \author{Florian Staub}\email{fnstaub@th.physik.uni-bonn.de}
 \affiliation{\AddrBonn}

 \author{Lorenzo Ubaldi}\email{ubaldi@th.physik.uni-bonn.de}
 \affiliation{\AddrBonn}

% \pacs{14.60.Pq, 12.60.Jv, 14.80.Cp}

\begin{abstract}
The distinguishing feature of the Dine-Fischler-Srednicki-Zhitnitsky (DFSZ) axion is that it couples to the electroweak Higgs fields. There
is thus an immediate connection between the Peccei-Quinn (PQ) scale and the weak scale. We wish to 
incorporate the DFSZ axion in a complete supersymmetric model, valid at all scales, and then 
to implement it in a numerical code connecting the high scale and the low scale physics on a quantitative 
level. We find that the simplest supersymmetric DFSZ model, as proposed by Rajagopal et al. in 1990, is inconsistent 
when we consider the minimization of the scalar potential. The problem is that we obtain a negative 
squared mass for the saxion, the scalar partner of the axion, at the minimum. We then consider the 
minimal extension in order to get a consistent model for all scales: one has to include an additional 
explicit sector to spontaneously break the PQ symmetry. In the complete model we can
determine the mass of the axino, the fermionic partner of the axion. It is useful to distinguish two cases: (1) the supersymmetry (SUSY) 
breaking scale is lower than the PQ breaking scale, and (2) the scales are comparable. We find that the axino
is very light in (1), while its mass is generically of the order 
of the other soft SUSY breaking masses in (2). We have implemented SUSY breaking via generic 
soft breaking terms, and thus make no explicit statement about the form and mediation of SUSY breaking. 
This complete model can then be incorporated in a numerical code connecting the two scales. We 
briefly discuss the renormalization group equations and the couplings of the axion to gluons and photons. 
\end{abstract}
\maketitle

\section{Introduction}
The Minimal Supersymmetric Standard Model (MSSM)~\cite{Martin:1997ns} is a complete model, which in principle should explain all phenomena from low--energy physics up to possible observations at the TeV scale. In its full 
version it contains over a 100 free parameters. To make it more predictive one typically makes simplifying assumptions at the unification scale reducing the number of parameters to only 5. This is for example the case 
in the constrained minimal supersymmetric model (CMSSM)~\cite{Drees:1992am}. The strength of such a model is that one can test it in a self consistent way against basically all experimental data. This includes a vast number of experiments in a wide energy range, for example measurements of the anomalous magnetic moment of the muon, LHC data, as well as direct and indirect searches for dark matter. This is possible thanks to the renormalization group equations (RGEs) which relate parameters at different energy scales~\cite{Martin:1993zk, Jack:1994kd}. Because of its simplicity and its predictive power the CMSSM has been widely studied for years. Due to the recent strict lower mass bounds on superpartner masses found at the LHC the CMSSM is in some tension with the Higgs mass measurement, as well as fine-tuning requirements~\cite{Bechtle:2013mda, Buchmueller:2011ab}.

A theoretical problem which is left unaddressed in the MSSM and in the CMSSM is the strong CP problem~\cite{Peccei:1988ci}. 
The most plausible solution is provided by the axion~\cite{Weinberg:1977ma, Wilczek:1977pj}, which can be implemented in two ways. In the first, the axion couples to new heavy quarks~\cite{Kim:1979if, Shifman:1979if} charged under the color $SU(3)$ gauge group and the global Peccei--Quinn (PQ) $U(1)_{\rm PQ}$ symmetry~\cite{Peccei:1977hh}. Models of this kind are referred to as KSVZ. Here the axion sector is decoupled
from the low-energy fields and we do not consider these models any further. In the second implementation, 
the DFSZ model~\cite{Dine:1981rt, Zhitnitsky:1980tq}, the axion couples to two electroweak Higgs-doublet fields, 
which are charged under the $U(1)_{\rm PQ}$; the Standard Model fermions then also carry PQ charge.  Employing the axion solution in a supersymmetric model requires promoting the axion to a superfield, which introduces two other new particles: the saxion, a scalar, and the axino, a fermion. The axion and also the axino have the virtue of themselves being good dark matter candidates~\cite{Dine:1981rt, Rajagopal:1990yx}.

Indeed the axino as a dark matter candidate has been widely explored. However the studies 
in the literature~\cite{Covi:2001nw, Kim:2001sh, Covi:2009pq,  Choi:2011yf, Bae:2011iw, Bae:2011jb, Strumia:2010aa} are not based on a complete model. Instead they consider the axion/axino sector separately, 
represented by only a few operators, often in an effective field theory framework, which encode the interactions 
of matter with the axino relevant only to its cosmological and thermal history. In that way one can get constraints that are fairly model independent. However with this approach it is difficult to correlate constraints coming from different observables at different energies, in particular also relating to the electroweak sector. If one wanted 
to do so then one would have to refer to a specific more complete model. 

A step in the direction of a phenomenological axion model in supersymmetry (SUSY) to study constraints in a fashion 
similar to the CMSSM, was taken by Baer and collaborators~\cite{Baer:2008yd, Baer:2009ms, Baer:2009vr, 
Baer:2010wm, Baer:2011hx, Baer:2011eca}. In particular in a recent paper~\cite{Bae:2013hma} they study 
a SUSY DFSZ axion model. Their starting point consists of two different benchmark models initially {\it without} the 
axion: one is denoted ``radiatively-driven natural SUSY"~\cite{Baer:2012up}, the other is the minimal 
supergravity model (mSUGRA). They first generate the (non-axion/axino) SUSY model mass spectra with ISAJET~\cite{Paige:2003mg} for both models, and only as a second step, {\it after} running the RGEs, they 
add the axion superfield with its interactions and study the resulting phenomenology. Thus these interactions 
are treated separately and do not enter in the details of the numerical calculations of the spectrum generator. 
This is somewhat justified by the fact that the interactions of the axion supermultiplet with matter are 
suppressed by inverse powers of the high PQ scale, $f_a \sim 10^{12}$ GeV. However we believe that it 
would be desirable to go a step further, and include the axion from the start and study the full model 
consistently. By using the same boundary conditions at the GUT or SUSY breaking scale for the soft 
parameters in the axion/axino 
sector as for the MSSM the model will become more predictive.
This should also be particularly relevant when extending the model to R-parity violating supersymmetry, 
where the axino also mixes with the neutrinos and neutralinos~\cite{Kim:2001sh}. 

The aim of the current paper is to study what are the minimal ingredients and assumptions needed to define 
a self consistent SUSY axion model, in order to later study its phenomenological implications with numerical 
tools available today. We concentrate on a DFSZ, R--parity conserving SUSY model. As SUSY already 
requires two electroweak Higgs doublets, this model is more economical than its KSVZ counterpart. The 
minimal model one might first consider is the one introduced in Ref.~\cite{Rajagopal:1990yx}. We show that this model has an inconsistent scalar sector: the saxion becomes tachyonic. We are thus forced to add superpotential terms that spontaneously break the PQ 
symmetry. The resulting model is self consistent and predictive. One prediction, for example, is that the 
mass of the heavy Higgs is modified compared to the MSSM.
 
In the context of these models we give what we consider a sharper and more intuitive answer to a question 
which has been investigated in the literature~\cite{Tamvakis:1982mw,Chun:1992zk,Chun:1995hc}: what is the 
axino mass, $m_{\tilde a}$, with broken supersymmetry? We consider two cases. In the first and simpler case, the SUSY 
breaking scale, $M_{\rm SB}$, is much smaller than the PQ breaking scale, $f_a$. Then we find that $m_{\tilde 
a} \sim \mathcal{O}\left( \frac{M^2_{\rm SUSY}}{f_a} \right)$, with $M_{\rm SUSY} \sim$ TeV the scale of the 
soft SUSY breaking terms. So the axino is very light. In the second case, we consider $M_{\rm SB} \geq f_a$, 
which is representative of gravity mediated SUSY breaking models where typically $M_{\rm SB} \sim M_{\rm GUT}$,
 and find that $m_{\tilde a} \sim M_{\rm SUSY}$. Our computations agree with previous claims while providing a hopefully new and edifying perspective
on the issue.

\section{The minimal inconsistent model}

It is not straightforward to embed the DFSZ axion in supersymmetric
models. The trouble is that the non--supersymmetric DFSZ
model~\cite{Dine:1981rt, Zhitnitsky:1980tq} contains a term $g
\varphi^2 H_u^\dagger H_d$ in the scalar potential, where the phase of
the complex scalar field $\varphi$ is the axion, $H_{u,d}$ are
  two complex Higgs doublets, and $g$ is a dimensionless real coupling constant. In supersymmetry this term can 
not be obtained from a renormalizable superpotential. To circumvent this difficulty Rajagopal et al. proposed~\cite{Rajagopal:1990yx} a recipe which  consists
 in replacing the term $\mu \hat{H}_u \hat{H}_d$  in the superpotential (the hat here denotes a superfield) with the term
\begin{equation}
\label{eq:effMuTerm}
 c_1 \hat{A} \hat{H}_u \hat{H}_d \ ,
\end{equation}
where $\hat A$ is the axion superfield and $c_1$ is a dimensionless constant. Once the scalar component of the superfield, $A$, 
gets a vacuum expectation value (VEV) $\langle A \rangle \sim f_a$, the parameter $\mu_{\rm eff}$ is given by $c_1 f_a$. This implies that $c_1$ has to be tuned to $10^{-9}$ or $10^{-10}$ in order to get $\mu_{\rm eff}$ at the weak scale. Despite being not a nice feature, this is 
no worse than the tuning which is needed for the coupling $g \leq 10^{-9}$ in the original DFSZ model~\cite{Dine:1981rt}, as already 
observed in Ref.~\cite{Rajagopal:1990yx}.

A more elegant implementation of the DFSZ axion in SUSY can be obtained by dropping the requirement of renormalizability and 
writing a higher dimensional operator of the form $g \frac{\hat A^2}{M_{\rm Pl}}  \hat H_u \hat H_d$, where $M_{\rm Pl}$ is the Planck mass. With a VEV $\langle A \rangle \sim 10^{11}$ GeV, this results in a $\mu_{\rm eff}$ at the correct scale and thus provides a solution to the $\mu$--problem~\cite{Kim:1983dt}.

Nevertheless a renormalizable operator is easier to implement in a numerical study of the complete model, which is our main goal. 
Thus we choose the operator of Eq.~\eqref{eq:effMuTerm} and accept the tuning of $\mu_{\rm eff}$ for the time being. A further justification for the choice of this operator comes from a recent study~\cite{Honecker:2013mya}, where the authors
have derived a consistent realization of a SUSY DFSZ axion model within String
Theory which can provide a microscopic explanation for the simple recipe of Rajagopal et al.

Let us then consider the model defined by the MSSM with the $\mu$--term replaced by Eq.~\eqref{eq:effMuTerm}.
The full superpotential, assuming R-parity conservation, reads
\begin{equation} \label{eq:W}
W = Y_u \hat{U} \hat{Q} \hat{H}_u + Y_d \hat{D} \hat{Q} \hat{H}_d + Y_e \hat{E} \hat{L} \hat{H}_d +  c_1 \hat{A} 
\hat{H}_u \hat{H}_d \ .
\end{equation}
$Y_{u,d,e}$ are $3\times 3$ Yukawa matrices. We have suppressed generation
  and $SU(2)$ and $SU(3)$ gauge indices.  The superfield $\hat A$ and
the Standard Model superfields carry PQ charges, as is distinctive of
DFSZ axion models, such that each term is invariant under the global
$U(1)_{\rm PQ}$.
 We will see that the
physical axion is a linear combination of the CP odd scalar
components of $\hat A$, $\hat H_u$ and $\hat H_d$, in complete analogy
with the non-SUSY model of Ref.~\cite{Dine:1981rt}. In order to
  compute the physical spectrum, one {\it assumes} that $A$ gets a vacuum
expectation value (VEV) $\langle A \rangle \sim f_a$. We show that simply assuming such a VEV
without specifying explicitly the PQ breaking mechanism and
  stabilizing the PQ potential leads to an inconsistency in the
model.

 The scalar potential is given by
\begin{align}
 V &=  V_F + V_D + V_{\rm soft} \, ,
\end{align}
where
\begin{align}
V_F =& \sum_\phi\left|\frac{\partial \widetilde{W}(\phi)}{\partial \phi}\right|^2  \, , \\[10 pt]
V_D =& \frac{1}{2} g_i^2 \left(\Phi T_\Phi^{i,a} \Phi^*\right)\left(\Phi T_\Phi^{i,a} \Phi^*\right) \, , \\[10 pt]
V_{\rm soft} = & T_u \tilde{u} \tilde{q} H_u + T_d \tilde{d} \tilde{q} H_d + T_e \tilde{e} \tilde{l} H_d + T_{c1} A H_u H_d  
\nonumber\\
 &  + m^2_a |A|^2 + m^2_{H_u} |H_u|^2 + m^2_{H_d} |H_d|^2+ \tilde{\phi}^\dagger m^2_{\tilde{\phi}} \tilde{\phi}  
\label{eq:soft1} \, ,
\end{align}
with $\tilde{\phi} \ni \{\tilde{e},
\tilde{l},\tilde{d},\tilde{u},\tilde{q} \}$, as well as
  $A,\,H_u,\,H_d$ the scalar components of the respective superfields. Here
  $\widetilde W(\phi)$ denotes the superpotential evaluated as a
  function of scalar fields. $T_{u,d,e,c1}$ are the trilinear soft
  breaking terms \cite{Nilles:1982dy}, elsewhere often denoted
  $A$. $T_\Phi^{i,a}$ are the gauge generators.  To avoid clutter we
take the soft parameters to be real in the following equations.  This
restriction does not affect our conclusions. The
conventional $B\mu$--term resulting from
Eq.~\eqref{eq:soft1} is given by $B_{\rm eff} = T_{c1} \langle A
\rangle$. 

The parameters have to fulfill the following tadpole equations
to minimize the scalar potential at tree-level
\begin{align}
\left.\frac{\partial V}{\partial \phi_d}\right|_{\phi=\sigma=0} = & m_{H_d}^2 v_d + \frac{1}{8} \Big[ 4 c_1^2 v_d 
\left(v_u^2+f_a^2\right)- 4 \sqrt{2} 
v_u B_{\rm eff} \nonumber  \\
& +v_d \left(g_1^2+g_2^2\right) (v^2_d-v^2_u)  \Big] = 0  \, , \label{eq:Atad1}\\ 
\left.\frac{\partial V}{\partial \phi_u}\right|_{\phi=\sigma=0}  = &  m_{H_u}^2 v_u + \frac{1}{8} \Big[ 4 c_1^2 v_u 
\left(v_d^2+f_a^2\right) -4 \sqrt{2} 
v_d B_{\rm eff} \nonumber \\& -v_u\left(g_1^2+g_2^2\right)(v^2_d-v^2_u)  \Big] = 0 \, , \label{eq:Atad2}\\
\left.\frac{\partial V}{\partial \phi_a} \right|_{\phi=\sigma=0} = & f_a m_a^2 + \frac{1}{2 f_a} \Big[ \mu_{\rm eff}^2
\left(v_d^2+v_u^2\right)\Big. \nonumber \\
&\Big.-\sqrt{2} 
v_d v_u B_{\rm eff} \Big]=0 \ . \label{eq:Atad3}
\end{align}
We have parametrized the scalar fields as in Ref.~\cite{Dine:1981rt}:
\begin{eqnarray}
&H_d = \frac{1}{\sqrt{2}}\left(\phi_d + i \sigma_d + v_d\right)\,,\ \ H_u = \frac{1}{\sqrt{2}}\left(\phi_u + i \sigma_u + v_u
\right) & \nonumber \\
&A = \frac{1}{\sqrt{2}}\left(\phi_a + i \sigma_a + f_a\right) \ .  &  \label{eq:VEV1}
\end{eqnarray}
The derivatives in Eqs.~(\ref{eq:Atad1})-(\ref{eq:Atad3}) are evaluated at
  the minimum, where $\phi_{d,u,a}=\sigma_{d,u,a}=0$. All tadpole equations
  and mass matrices have been calculated with the 
  public code {\tt SARAH} \cite{Staub:2008uz,Staub:2009bi,Staub:2010jh,Staub:2012pb,Staub:2013tta}.

Upon closer examination, Eq.~\eqref{eq:Atad3} presents a problem.  In order
  to solve the hierarchy problem, the scale of the soft SUSY breaking terms~\footnote{
  $M_{\rm SUSY}$ is the scale of the soft terms and is typically in the TeV range. It should
  not be confused with the SUSY breaking scale $M_{\rm SB}$ which can be much higher, depending
  on the mechanism that mediates the SUSY breaking. For instance, one has $M_{\rm SB}\sim M_{\rm GUT}$ in gravity mediation,
  and $M_{\rm SB} \sim M_{\rm messenger}$ in gauge mediation, where $M_{\rm messenger}$ is the messenger scale.}, $M_{\rm
    SUSY}$, should be of order $M_W$. One would expect that $m_a \sim M_{\rm SUSY}$. For proper electroweak
  symmetry breaking, we must also have $\mu_{\mathrm{eff}}^2,\,B_{\mathrm{eff}}=
  \mathcal{O}(M_W^2)$. Under these conditions Eq.~\eqref{eq:Atad3} is not
  soluble.  Let us then fix $\mu_{\rm eff}^2$ and $B_{\rm eff}$ at $M_{\rm
    SUSY}^2$ and solve for $m_{a}$. Then $m_a \sim M_{\rm SUSY}^2 / f_a$ is
  tiny.  This has an important consequence: it leads to a negative squared mass
eigenvalue for the scalar field that we can identify as the saxion.

Before we show this let us briefly compute the CP odd scalar sector. After
replacing the soft mass terms with the solutions of the tadpole equations the
mass matrix squared in the basis $(\sigma_d, \sigma_u, \sigma_a)$ reads in the
Landau gauge
\begin{equation}
 \mathcal{M}^2_{\mathrm{CP\,odd}}=\left(
\begin{array}{ccc}
 B_{\rm eff} t_\beta  & B_{\rm eff} & \frac{B_{\rm eff} t_\beta v}{\sqrt{t_\beta^2+1} f_a} \\
 B_{\rm eff} & \frac{B_{\rm eff}}{t_\beta} & \frac{B_{\rm eff} v}{\sqrt{t_\beta^2+1} f_a} \\
 \frac{B_{\rm eff} t_\beta v}{\sqrt{t_\beta^2+1} f_a} & \frac{B_{\rm eff} v}{\sqrt{t_\beta^2+1} f_a} & \frac{B_{\rm eff} t_
\beta
   v^2}{\left(t_\beta^2+1\right) f_a^2}
\end{array}
\right) \, .\label{CPodd}
\end{equation}
Here we have written $t_\beta\equiv\tan\beta\equiv\frac{v_u}{v_d}$ for the
  ratio of the vacuum expectation values. The matrix Eq.~(\ref{CPodd})
  has two eigenvalues which are exactly zero. One is associated with the
  Goldstone boson which gets absorbed by the massive $Z$ boson. The other is
  associated with the axion, the Goldstone boson of the spontaneously broken
  (global) PQ symmetry. This represents a check that the SUSY breaking effects
have not spoiled the Goldstone theorem \cite{Goldstone:1961eq}.  The third
eigenvalue is the mass squared of the physical CP odd Higgs boson 
\begin{equation}
m_A^2 = B_{\rm eff} \left[t_\beta+\frac{1}{t_\beta}+ \frac{t_\beta
    \,v^2}{\left(t_\beta^2+1\right) f_a^2}\right] \, ,
\end{equation}
which is the same as in the MSSM, apart from the very small correction given by the last term.

Let us turn now to the scalar mass matrix squared for the CP even states. After rotating the Higgs
fields $(\phi_d, \phi_u) \to (h,H)$, it reads in the basis $(h,H,\phi_a)$
\begin{widetext}
\begin{equation}
 \mathcal{M}^2_{\mathrm{CP\,even}}=
\left(
\begin{array}{ccc}
\frac{\left[16 \mu_{\rm eff}^2 t_\beta^2+f_a^2 \left(g_1^2+g_2^2\right) \left(t_\beta^2-1\right)^2\right]v^2}{4
   f_a^2 \left(t_\beta^2+1\right)^2} & \frac{\left[f_a^2 \left(g_1^2+g_2^2\right)-4 \mu_{\rm eff}^2\right]
   t_\beta \left(t_\beta^2-1\right) v^2}{2 f_a^2 \left(t_\beta^2+1\right)^2} & \frac{2 \left[\mu_{\rm eff}^2
   \left(t_\beta^2+1\right)-B_{\rm eff} t_\beta\right] v}{f_a \left(t_\beta^2+1\right)} \\[3mm]
 \frac{\left[f_a^2 \left(g_1^2+g_2^2\right)-4 \mu_{\rm eff}^2\right] t_\beta \left(t_\beta^2-1\right) v^2}{2
   f_a^2 \left(t_\beta^2+1\right)^2} & \frac{\left[f_a^2 \left(g_1^2+g_2^2\right)-4 \mu_{\rm eff}^2\right] v^2
   t_\beta^3+B_{\rm eff} f_a^2 \left(t_\beta^2+1\right)^3}{f_a^2 t_\beta \left(t_\beta^2+1\right)^2} &
   \frac{B_{\rm eff} \left(t_\beta^2-1\right) v}{f_a \left(t_\beta^2+1\right)} \\[3mm]
 \frac{2 \left[\mu_{\rm eff}^2 \left(t_\beta^2+1\right)-B_{\rm eff} t_\beta\right] v}{f_a \left(t_\beta^2+1\right)} &
   \frac{B_{\rm eff} \left(t_\beta^2-1\right) v}{f_a \left(t_\beta^2+1\right)} & \frac{B_{\rm eff} t_\beta v^2}{f_a^2
   \left(t_\beta^2+1\right)} \\
\end{array}
\right) \ .
\end{equation}
\end{widetext}
Neglecting the entries with a $v^2/f_a^2$ suppression and approximating $t_\beta^2+1=t_\beta^2-1=t_\beta^2$ this 
matrix has the form
\begin{equation}
\left(
\begin{array}{ccc}
 \frac{1}{4} \left(g_1^2+g_2^2\right) v^2 & \frac{\left(g_1^2+g_2^2\right) v^2}{2 t_\beta} & \frac{2 \left(\mu_{\rm eff}^2
   t_\beta-B_{\rm eff}\right) v}{f_a t_\beta} \\[3mm]
 \frac{\left(g_1^2+g_2^2\right) v^2}{2 t_\beta} & \frac{\left(g_1^2+g_2^2\right) v^2}{t_\beta^2}+B_{\rm eff} t_\beta &
   \frac{B_{\rm eff} \,v}{f_a} \\[3mm]
 \frac{2 \left(\mu_{\rm eff}^2 t_\beta-B_{\rm eff}\right) v}{f_a t_\beta} & \frac{B_{\rm eff} \,v}{f_a} & 0
\end{array} 
\right) \, .
\end{equation}
The determinant is given by 
\begin{align}
&-\frac{1}{4 f_a^2 t_\beta^4}\Big\{v^4 \left(g_1^2+g_2^2\right) \left[B_{\rm eff} \left(t_\beta^2+4\right)-4 \mu_{\rm eff}
^2 t_\beta\right]^2 \nonumber \\
&\hspace{1cm}+16 B_{\rm eff}
   t_\beta^3 v^2 \left(B_{\rm eff}-\mu_{\rm eff}^2 t_\beta\right)^2\Big\} \, .
\end{align}
$B_{\rm eff}$ must be positive, otherwise the mass of the charged Higgs is
below the $W$ boson mass.  Hence the determinant is always negative and the
saxion is a tachyon. This is a new result to the best of our knowledge. We conclude that this model is not consistent.  As the
issue can be traced back to the minimization condition, Eq.~\eqref{eq:Atad3},
corresponding to the PQ-breaking VEV, the problem can be fixed by adding terms in
the superpotential to spontaneously break the PQ symmetry and stabilize the
  PQ breaking scale.

%%%%%%%%%%%%%%%%%%%%%%%%%%%%%%%%%%%%%%%%%%%%%%%%%%%%
%%%%%%%%%%%%%%%%%%%%%%%%%%%%%%%%%%%%%%%%%%%%%%%%%%%%
%%%%%%%%%%%%%%%%%%%%%%%%%%%%%%%%%%%%%%%%%%%%%%%%%%%%

\section{A self-consistent model} \label{sec:consistent}

We add the following terms~\cite{Kim:1983ia} to the superpotential in Eq.~\eqref{eq:W} 
\begin{equation} \label{eq:WPQ}
W_{\rm PQ} =  \lambda \hat{\chi} \left(\hat{A} \hat{\bar{A}}  - \frac{1}{4} f_a^2\right) \ ,
\end{equation}
with the distinct superfields $\hat{A},\,\hat{\bar{A}}$, as well as
  $\hat\chi$.  $\hat{\bar A}$ carries a PQ charge opposite to $\hat A$, while
$\hat \chi$ is PQ neutral.  Assuming that a global R symmetry~\footnote{For a gauged R-symmetry see Ref.~\cite{Chamseddine:1995gb}.} 
forbids terms quadratic
and cubic in $\hat \chi$, we have written all the terms consistent with the
gauge and PQ symmetries, as well as with R-parity. 

Considering the superpotential $W_{\rm PQ}$ alone with unbroken SUSY, one has two heavy superfields,
$\hat \chi$ and $\frac{1}{\sqrt 2} (\hat A + \hat{\bar A})$, with masses $\sim f_a$, and a massless superfield, 
$\frac{1}{\sqrt 2} (\hat A - \hat{\bar A})$. The latter can be identified as the axion superfield. If SUSY is broken at a scale
$M_{\rm SB} \ll f_a$ we can integrate out the heavy superfield in a supersymmetric fashion and then consider the SUSY
breaking effects in the resulting effective theory. We will study this case in the next section. When $M_{\rm SB} \gg f_a$, like 
in supergravity models, we have to take into account the SUSY breaking effects also for the fields with masses $\sim f_a$.
This is the case we consider in the rest of this section.

After electroweak symmetry
breaking (EWSB) $\chi$ gets a VEV, $v_\chi$, thus the R symmetry is broken. The corresponding R-axion has a 
mass 
of order $M_{\rm SUSY}$ because the R-symmetry is also explicitly broken by the soft terms.
  Beyond those in Eq.~\eqref{eq:soft1} we have the soft-breaking terms
\begin{equation} \label{eq:soft2}
 V_{\mathrm{soft}}^{\rm PQ}=T_\lambda \chi A \bar{A} - L_V \chi + m^2_{\bar{a}} |\bar{A}|^2 + m^2_{\chi} |\chi|^2 \ .
\end{equation}
The trilinear and linear terms, with coefficients $T_\lambda$ and $L_V$, will play an important
role when we discuss the mass of the axino. Note that one expects $L_V \sim
M_{\rm SUSY} f_a^2$. 

After PQ and EW breaking the fields $H_d$, $H_u$, $A$, $\bar{A}$ and $\chi$ receive VEVs:
\begin{eqnarray}
&A = \frac{1}{\sqrt{2}}\left(\phi_a + i \sigma_a + v_a\right)\,,\ \ \bar{A} = \frac{1}{\sqrt{2}}\left(\phi_{\bar{a}} + i 
\sigma_{\bar{a}} + v_{\bar{a}}\right)& \nonumber \\
&\chi = \frac{1}{\sqrt{2}}\left(\phi_\chi + i \sigma_\chi + v_\chi \right), &
\end{eqnarray}
with $v_a v_{\bar a} = \frac{1}{2} f_a^2$. 
The fields $H_u,H_d$ are parametrized as in Eq.~\eqref{eq:VEV1}. 
The tadpole equations for $\phi_a, \phi_{\bar a}$ and $\chi$ read
\begin{align}
\label{eq:tadVx}
\frac{\partial V}{\partial \phi_a} &= m_a^2 v_a+\frac{1}{4} \Big[2 v_a \left(c_1^2 \left(v_d^2+v_u^2\right)\right) 
\nonumber \\
&  +2 \sqrt{2} (v_\chi v_{\bar{a}} T_\lambda  -v_d v_u T_{c1})+ 2 \lambda ^2 v_a  v_\chi^2 \Big] = 0 \, , \\
\label{eq:tadVxb}
\frac{\partial V}{\partial \phi_{\bar{a}}} &= m_{\bar{a}}^2 v_{\bar{a}}+\frac{1}{4} \Big(  -2 \lambda c_1 v_\chi v_d v_u  
+2 
\lambda^2 v_{\bar{a}}  
   v_\chi^2  \nonumber \\
   &  +2 \sqrt{2} v_\chi v_a T_\lambda \Big) =0 \, , \\
\label{eq:tadChi}   
\frac{\partial V}{\partial \phi_{\chi}} &=  m^2_{\chi} v_\chi - \sqrt{2} L_V+\frac{1}{2} \Big[- \lambda c_1 v_d v_u v_{\bar{a}} 
   +\sqrt{2} v_a v_{\bar{a}} T_\lambda  \nonumber \\
   &  +v_\chi \lambda ^2 \left(v_a^2+v_{\bar{a}}^2\right)\Big]  = 0 \, .
\end{align}
We can consistently solve these equations keeping all the soft parameters at the $M_{\rm SUSY}$ 
scale. In particular we can solve the last equation for $v_\chi$ and find
 \begin{equation} \label{eq:vchi}
 v_\chi = \frac{2\sqrt{2}}{\lambda^2 (v_a^2 + v_{\bar a}^2)}L_V -\frac{\sqrt{2} v_a v_{\bar a}}{\lambda^2 (v_a^2 + 
 v_{\bar a}^2)} T_\lambda + \mathcal{O}\left(\frac{M^2_{\rm SUSY}}{f_a} \right)  \ .
 \end{equation}
%\FS{
%\begin{equation} \label{eq:vchi}
%v_\chi = \sqrt{\frac{2 (m^2_{\bar a} v^2_{\bar a} - m^2_{a} v_a^2 )}{\lambda(v_a^2 - v_{\bar a}^2)}}+ \mathcal{O}\left(\frac{M^2_{\rm SUSY}}{f_a} \right)  \ .
%\end{equation}
%}

This result will be important for the discussion below on the axino mass. 

First we comment on the scalar masses in our model. Eq.~(\ref{eq:effMuTerm}) introduces a mixing between 
the MSSM Higgs sector and the axion sector. It turns out that the correction to the light Higgs mass, $m_h$, is of 
order $
\mu_{\rm eff}^2/f_a$, which is negligible. As a consequence the usual upper limit $m_h < m_Z$ holds at tree level. 
On 
the other hand the tree-level mass of the heavy Higgs, $m_H$, is modified as
\begin{align} \label{eq:Hmass}
m^2_H = \frac{\left(2 B_{\rm eff}+\sqrt{2} \mu_{\rm eff}  v_\chi \lambda \right)}{\sin(2 \beta ) } +\frac{v^2}{4} \sin ^2(2 
\beta )\left(g_1^2+g_2^2\right) \, .
\end{align}
Due to $v_\chi \neq 0$ this can be potentially different from the MSSM. If we neglect the small mixing between the MSSM and the 
axion sector the three squared mass eigenvalues stemming from the mixing among $(\phi_a, \phi_{\bar{a}}, \phi_
\chi)
$ are given by 
\begin{equation}
\label{eq:saxionMass}
\frac{4 L_V T_\lambda+T_\lambda^2 f_a^2}{f_a^2 \lambda ^2}\,,\ \ \frac{f_a^3 \lambda ^2\pm4 \sqrt{2} L_V}{2
   f_a} \, .
\end{equation}   
The first is the smaller one and is associated it with the saxion mass squared.
In models where the SUSY breaking effects are mediated by gravity, one expects the linear and trilinear soft terms to be 
of order $M_{\rm SUSY} \sim m_{3/2}$, with $m_{3/2}$ the gravitino mass. In such a scenario the saxion mass is then of order $M_{\rm SUSY}$.
The other two scalars have a mass of order $f_a$.
In the scalar CP-odd sector we find a massless axion~\footnote{We are neglecting here QCD instanton effects that 
generate a small axion mass.} as expected.

The extended neutralino mass matrix reads, in the basis $\left(\lambda_{\tilde{B}}, \tilde{W}^0, \tilde{H}_u^0, \tilde{H}_d^0, 
\tilde{A}, \tilde{\bar{A}}, \tilde{\chi}\right)$
\begin{widetext}
\begin{equation} 
m_{\chi^0} = \left( 
\begin{array}{ccccccc}
M_1 &0 &\frac{1}{2} g_1 v_u  &-\frac{1}{2} g_1 v_d  &0 &0 &0\\ 
0 &M_2 &-\frac{1}{2} g_2 v_u  &\frac{1}{2} g_2 v_d  &0 &0 &0\\ 
\frac{1}{2} g_1 v_u  &-\frac{1}{2} g_2 v_u  &0 &- \frac{1}{\sqrt{2}} c_1 v_a  &- \frac{1}{\sqrt{2}} c_1 v_d  &0 &0\\ 
-\frac{1}{2} g_1 v_d  &\frac{1}{2} g_2 v_d  &- \frac{1}{\sqrt{2}} c_1 v_a  &0 &- \frac{1}{\sqrt{2}} c_1 v_u  &0 &0\\ 
0 &0 &- \frac{1}{\sqrt{2}} c_1 v_d  &- \frac{1}{\sqrt{2}} c_1 v_u  &0 &\frac{1}{\sqrt{2}} v_{\chi} \lambda  &\frac{1}
{\sqrt{2}} v_{\bar{a}} \lambda \\ 
0 &0 &0 &0 &\frac{1}{\sqrt{2}} v_{\chi} \lambda  &0 &\frac{1}{\sqrt{2}} v_a \lambda \\ 
0 &0 &0 &0 &\frac{1}{\sqrt{2}} v_{\bar{a}} \lambda  &\frac{1}{\sqrt{2}} v_a \lambda  &0\end{array} 
\right) \, .
 \end{equation} 
\end{widetext} 
In the limit $v_{a} = v_{\bar{a}} = \frac{f_a}{\sqrt{2}}$, $c_1 v_u \to 0$, $c_1 v_d \to 0$ the lower right $3\times3$ 
block has 
the singular values~\cite{Dreiner:2008tw}  
\begin{equation} \label{eq:max}
- \frac{1}{\sqrt{2}} v_{\chi} \lambda \,, \ \  \frac{1}{2\sqrt{2}} \Big(\pm \sqrt{v_\chi^2+4 f_a^2} \lambda  + 
v_{\chi} \lambda \Big) \, .
\end{equation}
The first is associated with the phyiscal axino. Its mass is proportional to $v_\chi$, therefore of order $M_{\rm SUSY}$.
It is important to notice that the fact that the field $\chi$ develops a VEV $v_\chi$ is tightly connected to the SUSY breaking
effects. If the SUSY breaking scale is much lower than the PQ scale, we can integrate out the heavy superfield $\hat \chi$ supersymmetrically,
after which there would be no notion of $v_\chi$ any longer. When we study this limit in the next section we will see that the resulting 
axino is very light.

\begin{figure}[hbt]
\includegraphics[width=\linewidth]{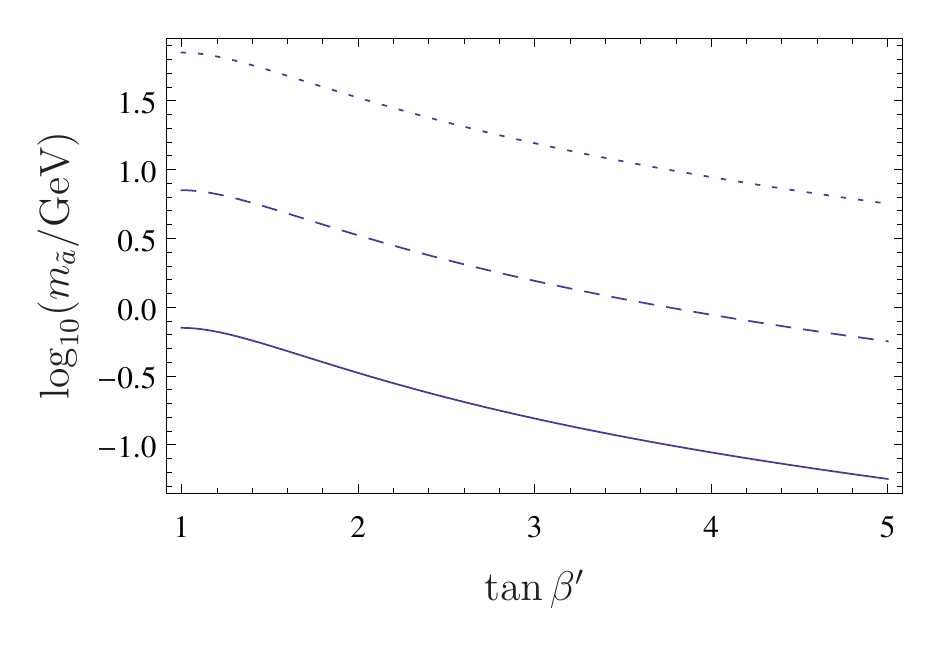} 
\caption{The mass of the axino as a function of $\tan\beta' = \sqrt{v_{\bar a} / v_a}$. We fixed 
$f_a = 10^{12}$~GeV, $\lambda=10^{-2}$, $T_\lambda=-1$~TeV and used $v_\chi = 10^4$~GeV (dotted line), 
$v_\chi = 10^3$~GeV (dashed line), $v_\chi = 10^2$~GeV (plain line).}
\label{fig:AxinoTBp}
\end{figure}

In the model considered in this section we also have an extra handle on the axino mass. We can relax the assumption $v_{a} = v_{\bar{a}} = 
\frac{f_a}{\sqrt{2}}$ and consider a hierarchy between the two VEVs, for example $v_{\bar a} \gg v_a$. If we do so 
we 
find that the axino mass becomes lighter. In the limit $v_a \to 0$, keeping fixed $v_a v_{\bar a } = 1/2 f_a^2$, the 
axino mass tends to zero. We show the axino mass as a function of $\tan\beta' = \sqrt{v_{\bar a} / v_a}$ in 
Fig.~\ref{fig:AxinoTBp}. 

\subsection{Renormalization Group Equations}
% \section{Constrained setup and RGE running}

The full model defined by the superpotential terms in Eqs.~\eqref{eq:W} and \eqref{eq:WPQ} can be defined as the 
minimal 
supersymmetric DFSZ axion model. It is possible at this point to make some simplifying assumptions in a CMSSM 
fashion, 
and study phenomenology and constraints with a relatively small parameter space. 
This will be studied in a future work in detail but here we briefly comment on the main aspects.

In the Appendix we give the one-loop RGEs for the axino sector and the changes in 
the $\beta$ functions. As already stated in the introduction, the $\beta$ functions of 
the MSSM parameters change only by terms of order $\mu/f_a$, which is usually a negligible effect. 
This is in contrast to the NMSSM, where the coupling between the Higgs doublets and the gauge singlet,
the analogous of our $c_1$, can be of order one and can significantly change the MSSM RGEs.

In the CMSSM one uses universal boundary conditions. In the same spirit, the simplest choice to keep
the number of parameters to a minimum is to use the same boundary conditions also for the 
parameters in our new sector. Thus, our additional soft terms at the GUT scale are fixed by two parameters, $m_0^2$ and $A_0$:
\begin{eqnarray*}
& m_{a}^2 = m_{\bar a}^2 = m_{\chi}^2 \equiv m_0^2 \, , &  \\
& T_\lambda \equiv A_0 \lambda \,, \hspace{1cm} T_{c1} \equiv A_0 c_1 \, ,&
\end{eqnarray*}
while $L_V$ can be eliminated by using the minimization conditions of the vacuum. 
With this setup, we have immediately a lower limit on $m_0$ of several hundreds of GeV coming from squark searches at the LHC. If we neglect the small contributions 
from $c_1$ and $ T_{c_1}$, the relation $m_a^2 = m_{\tilde a}^2 = m_{\chi^2} \equiv m^2$ holds at each scale and 
we have effectively the three RGEs
\begin{align}
16 \pi^2 \frac{d}{dt} {m^2} =&  6 m^2 |\lambda|^2  + 2 |T_\lambda|^2 \, , \\
16 \pi^2 \frac{d}{dt} {\lambda} = & 3 \lambda |\lambda|^2 \, , \\
16 \pi^2 \frac{d}{dt} {T_{\lambda}} = & 9 T_\lambda |\lambda|^2  \, ,
\end{align}
which can even be solved analytically:
\begin{align}
\lambda(t) =& \frac{2 \pi }{\sqrt{\frac{3 (t_{GUT}-t)}{2}+\frac{4 \pi ^2}{\lambda ^2}}} \, , \\
T_\lambda(t) =& -\frac{16 i \sqrt{2} \pi ^3 A_0}{\lambda ^2 \left(3 (t-t_{GUT})-\frac{8 \pi ^2}{\lambda ^2}\right)^{3/2}} \, , \\
m^2(t) =& \frac{64 \pi ^4 m_0^2-8 \pi ^2 \lambda ^2 \left(A_0^2-3 m_0^2\right) (t_{GUT}-t)}{\left(3 \lambda ^2 (t_{GUT}-t)+8 \pi \
^2\right)^2} \, .
\end{align}
Here $t$ is the renormalization scale and $t_{GUT}$ the scale of grand unification where the boundary conditions have been applied.

%%%%%%%%%%%%%%%%%%%%%%%%%%%%%%%%%%%%%%%%%%%%%%%%%%
%%%%%%%%%%%%%%%%%%%%%%%%%%%%%%%%%%%%%%%%%%%%%%%%%%
\section{The axino mass}

In the above discussion we have parametrized the SUSY breaking effects in the soft terms and assumed a high SUSY breaking scale, $M_{\rm SB} \gg f_a$.
We have also seen that we have some heavy fields in the spectrum, with masses of order $f_a$. In this section we study how
integrating out the heavy fields affects the mass of the axino in the low energy theory. We distinguish two cases.
In the first we consider $M_{\rm SB} \gg f_a$, as in the previous section. Here we find that the axino mass remains of order $M_{\rm SUSY}$.
In the second we take the opposite limit, $M_{\rm SB} \ll f_a$, and find that the resulting axino is much lighter.

%We now consider the effects of $F$--term SUSY breaking in two cases: (a) when $\sqrt{F} \ll f_a$, and (b) when 
% $\sqrt{F} \sim f_a$. Case (a) is representative of models with global SUSY ({\it e.g.} gauge mediation), 
% while case (b) is 
%typical of supergravity.}

\subsection{High scale SUSY breaking}

\begin{figure}[hbt]
\includegraphics[width=0.5\linewidth]{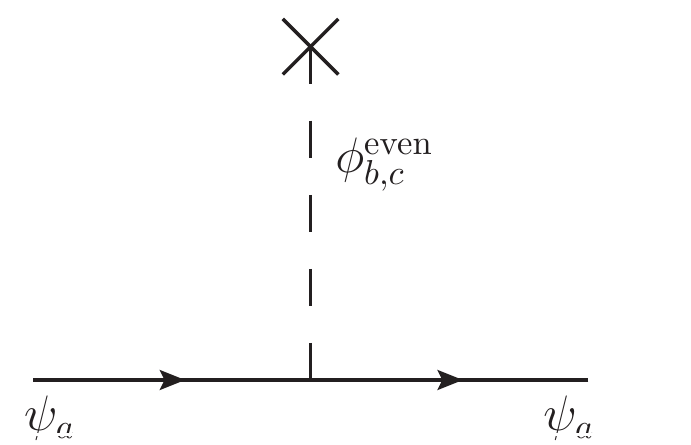} 
\caption{Diagrams contributing to the axino mass when integrating out the heavy fields $\phi_{b,c}$.}
\label{fig:intout}
\end{figure}

  We have seen in Eq.~\eqref{eq:max} that retaining all the fields we obtain an axino mass of order $M_{\rm SUSY}$.
  One may wonder what happens to the axino mass in the low-energy theory if we integrate out the heavy fields in 
  this scenario. We have to integrate them out component by component as SUSY is already broken. First we 
  diagonalize the scalar and fermionic mass matrices. As we have seen in eq.~\eqref{eq:saxionMass}, it is easy to 
  identify in the CP-even sector the light state, the saxion, which we denote $\phi_a^{\rm even}$ here, and the two heavy scalar states,
   with masses of order $f_a$, which we denote $\phi_b^{\rm even}$ and $\phi_c^{\rm even}$. 
   In the fermionic sector we have the axino,
  $\psi_a$, associated with the first eigenvalue in eq.~\eqref{eq:max}, and two heavy fermions, $\psi_b$ and $\psi_c$, associated with  
  the other two eigenvalues. At tree level the only contributions to the axino mass one can have when integrating out the heavy fields is depicted
  in Fig.~\ref{fig:intout}. Note that only the CP-even scalars contribute. It is easy to check that the yukawa coupling between the scalar and the two $\psi_a$'s is the same for $\phi_b^{\rm even}$ and $\phi_c^{\rm even}$. The scalar propagator ends in a tadpole. This is the key point. 
  The tadpole is given by $\frac{\partial V}{\partial \phi_i^{\rm even}}$, with $V$ the scalar potential. Then one is guaranteed that the tadpoles
  for $\phi_b^{\rm even}$ and $\phi_c^{\rm even}$ are zero, as this corresponds to the minimization condition of the scalar potential. In the 
  previous section we used $\frac{\partial V}{\partial \phi_i}=0$ for the gauge eigenstates, $\phi_i$. It is clear that the same condition holds for the mass 
  eigenstates $\phi_i^{\rm even}$ here, as we have $\frac{\partial V}{\partial \phi_i^{\rm even}} = \frac{\partial \phi_j}{\partial \phi_i^{\rm even}}\frac{\partial V}{\partial \phi_j} =0$. Thus the contribution of the diagrams in Fig.~\ref{fig:intout} vanishes and the axino mass at tree level does not
  change in the low energy. 
 One might worry about loop corrections. At worst these would be of order 
 $\frac{1}{16\pi^2} M_{\rm SUSY}$. Because of the $16 \pi^2$ loop-suppression they would not 
 provide any significant cancelation. We conclude that the axino mass in this scenario remains of order $M_{\rm SUSY}$.

%%%%%%%%%%%%%%%%%%%%%%%%%%%%%%%%%%%%%%%%%%%%%%%%%%%%
%%%%%%%%%%%%%%%%%%%%%%%%%%%%%%%%%%%%%%%%%%%%%%%%%%%%

%%%%%%%%%%%%%%%%%%%%%%%%%%%%%%%%%%%%%%%%%%%%%
%%%%%%%%%%%%%%%%%%%%%%%%%%%%%%%%%%%%%%%%%%%%%

\subsection{Low scale SUSY breaking}

If $M_{\rm SB} \ll f_a$ SUSY is still unbroken at the PQ scale and we can perform the following redefinitions of the 
superfields:
\begin{eqnarray}
\hat{\chi} & \to & \hat{\chi} \, , \\
\hat{A} & \to & \left( \frac{1}{2} f_a + \frac{1}{\sqrt{2}} \hat \Phi_H \right) e^{\sqrt{2} \frac{\hat \Phi_a}{f_a}} \, , \\
\hat{\bar A} & \to & \left( \frac{1}{2} f_a + \frac{1}{\sqrt{2}} \hat \Phi_H \right) e^{-\sqrt{2} \frac{\hat \Phi_a}{f_a}}  \, .
\end{eqnarray}
Inserting into Eq.~\eqref{eq:WPQ} we see that the superfields $\hat \chi$ and $\hat \Phi_H$ have masses of order 
$f_a$, while $\hat \Phi_a$ is massless. The latter is the axion superfield. This parametrization~\cite{Bae:2011jb} is 
useful because it explicitly shows that the original PQ transformation $\hat A \to e^{i\alpha} \hat A$ is now encoded 
in $\hat \Phi_a \to \hat \Phi_a + i \sqrt{2} \alpha f_a$. We recognize here the shift symmetry typical of axions that 
must be respected in the low energy theory. Let's consider the  superpotential in terms of the new superfields
\begin{align} \label{eq:Wrotated}
W_2 = & \frac{\lambda}{2} \hat\chi \hat\Phi_H (\hat\Phi_H + \sqrt{2} f_a ) \nonumber  \\
& + c_1 \left( \frac{1}{2} f_a + \frac{1}{\sqrt{2}} \hat \Phi_H \right) e^{\sqrt{2} \frac{\hat \Phi_a}{f_a}} \hat H_u \hat H_d.
\end{align}
We can integrate out the heavy fields in a supersymmetric fashion using their equations of motion: $\frac{\partial 
W_2}{\partial \hat\chi} = 0$ and $\frac{\partial W_2}{\partial \hat\Phi_H} = 0$. We find the following effective 
superpotential
\begin{equation} \label{eq:Weff}
W_{\rm eff} = \mu_{\rm eff} \hat H_u \hat H_d + \frac{c_1}{\sqrt{2}} \hat\Phi_a \hat H_u \hat H_d + \frac{c_1}{2} 
\sum_{n\geq 2} \frac{(\sqrt 2 \hat \Phi_a)^n}{f_a^{n-1}} \hat H_u \hat H_d.
\end{equation}
In the first term we have $\mu_{\rm eff} = \frac{c_1}{2} f_a$, while the last term contains higher dimension operators 
that we can safely neglect because they are suppressed by increasing negative powers of $f_a$. Next we 
consider the effects of SUSY breaking. The soft terms for the low energy field content read
\begin{align}
V_{\rm soft} = & T_u \tilde{u} \tilde{q} H_u + T_d \tilde{d} \tilde{q} H_d + T_e \tilde{e} \tilde{l} H_d + T_{c1} \Phi_a H_u 
H_d  
\nonumber\\
 & + B_{\rm eff} H_u H_d  + m^2_{H_u} |H_u|^2 + m^2_{H_d} |H_d|^2+ \tilde{\phi}^\dagger m^2_{\tilde{\phi}} 
\tilde{\phi}  \nonumber\\
 & + m^2_{a} (\Phi_a + \Phi_a^* )^2
\label{eq:soft2} \, .
\end{align}
The form of the last term is dictated by the shift symmetry, which would be otherwise violated if we wrote $m_a^2 |\Phi_a|^2$.
 Indeed this term in Eq.~\eqref{eq:soft2} gives a mass to the saxion, the real part of $\Phi_a$, but leaves 
the axion massless.
Note that after integrating out the heavy fields, the parameters $\lambda, T_\lambda, L_V$ do not appear any more
 in the low energy lagrangian.
 We parametrize the fields $H_u$ and $H_d$ as in Eq.~\eqref{eq:VEV1}, but do not assign a 
VEV to the field $\Phi_a$, as that would break the shift symmetry. Thus we write
$\Phi_a = \frac{1}{\sqrt{2}} (\phi_a + i \sigma_a)$. The tadpole equations read
 \begin{align}
\left.\frac{\partial V}{\partial \phi_d}\right|_{\phi=\sigma=0} = & m_{H_d}^2 v_d + \frac{1}{8} \Big[ 2 c_1^2 v_d  v_u^2 + 
8 \mu_{\rm eff}^2 v_d - 8 v_u B_{\rm eff} \nonumber  \\
& +v_d \left(g_1^2+g_2^2\right) (v^2_d-v^2_u)  \Big] = 0  \, , \label{eq:Atad4}\\ 
\left.\frac{\partial V}{\partial \phi_u}\right|_{\phi=\sigma=0}  = &  m_{H_u}^2 v_u + \frac{1}{8} \Big[ 2 c_1^2 v_u v_d^2  
+8 \mu_{\rm eff}^2 v_u -8 v_d B_{\rm eff} \nonumber \\
& -v_u\left(g_1^2+g_2^2\right)(v^2_d-v^2_u)  \Big] = 0 \, , \label{eq:Atad5}\\
\left.\frac{\partial V}{\partial \phi_a} \right|_{\phi=\sigma=0} = & v_d v_u T_{c1} - c_1 \mu_{\rm eff} (v_d^2 + v_u^2)=0 
\ . \label{eq:Atad6}
\end{align}
We see at this stage that the issue which made the model of the first section inconsistent is no longer present. 
The parameter $m_a$ is absent from these equations. Thus we can retain all the soft masses at the $M_{\rm SUSY}$
 scale. We find that the Higgs masses are now the same as in the MSSM, up to tiny corrections proportional to the small 
parameter $c_1$. The correction to the heavy Higgs mass in Eq.~\eqref{eq:Hmass} proportional to $v_\chi$ is no longer 
present, as $v_\chi$ is effectively zero in the limit $M_{\rm SB} \ll f_a$.
The axion is massless.  

The saxion mass comes almost entirely from the soft parameter $m_a$, as the contributions from the mixing with the Higgs fields
are suppressed by the tiny value of $c_1$, and it deserves a comment. If one considered minimal gauge 
mediation as an example of low scale SUSY breaking, then the parameter $m_a$ would only be generated at three loops, as $\Phi_a$ is
a gauge singlet. The saxion would then be very light and very problematic from the cosmological point of view~\cite{Banks:2002sd}. 
One could easily contemplate extended hidden sectors in the context of gauge mediation which would result in a heavier saxion~\cite{Carpenter:2009sw}.
Such extensions would likely produce a mass for the axino comparable to that of the saxion. However, it is our aim here to keep our model
minimal in order to make it as predictive as possible. Thus we do not consider any hidden sector or mediation mechanism but just parametrize the saxion mass as $m_a$.
This is a free parameter for us, which can be taken to be of order $M_{\rm SUSY}$, to avoid cosmological problems. A further clarification is then necessary. 
 In Ref.~\cite{Tamvakis:1982mw} the authors claimed that in theories with spontaneously broken SUSY 
with $M_{\rm SB} \ll f_a$ the saxion mass is at most $M_{\rm SUSY}^2 / f_a$. Their result relies on the assumption that 
the supertrace sum rule~\cite{Ferrara:1979wa} holds. The inclusion of the explicit soft SUSY breaking terms violates 
this assumption, and our saxion mass comes indeed from the soft term. Therefore our result is not in conflict with 
Ref.~\cite{Tamvakis:1982mw}.

Let us consider the fermions. In the basis $\left(\lambda_{\tilde{B}}, \tilde{W}^0, \tilde{H}_u^0, \tilde{H}_d^0, 
\tilde\Phi_a \right)$ the $5\times 5$ extended neutralino mass matrix reads
\begin{widetext}
\begin{equation} 
m_{\chi^0} = \left( 
\begin{array}{ccccccc}
M_1 &0 &\frac{1}{2} g_1 v_u  &-\frac{1}{2} g_1 v_d  &0 \\ 
0 &M_2 &-\frac{1}{2} g_2 v_u  &\frac{1}{2} g_2 v_d  &0 \\ 
\frac{1}{2} g_1 v_u  &-\frac{1}{2} g_2 v_u  &0 &- \mu_{\rm eff}  &- \frac{1}{2} c_1 v_d \\ 
-\frac{1}{2} g_1 v_d  &\frac{1}{2} g_2 v_d  &- \mu_{\rm eff}  &0 &- \frac{1}{2} c_1 v_u  \\ 
0 &0 &- \frac{1}{2} c_1 v_d  &- \frac{1}{2} c_1 v_u  &0  \end{array} 
\right) \ .
 \end{equation} 
\end{widetext} 
The smallest eigenvalue here is of order $c_1 v$, with $v$ of order the EWSB VEV, and it corresponds to the axino. 
Given that $c_1 = \frac{\mu_{\rm eff}}{f_a}$, the axino in this model has a mass of order $M^2_{\rm SUSY}/f_a \leq 
\mathcal{O}({\rm keV})$.

%%%%%%%%%%%%%%%%%%%%%%%%%%%%%%%%%%%%%%%%%%%%%
%%%%%%%%%%%%%%%%%%%%%%%%%%%%%%%%%%%%%%%%%%%%%

\subsection{Comments}
The axino mass has been widely discussed in the literature.
Tamvakis and Wyler~\cite{Tamvakis:1982mw} showed that in models with global SUSY the axino mass would be at 
most of order $\mathcal{O}\left(\frac{M^2_{\rm SUSY}}{f_a} \right)$ after SUSY breaking. Chun, Kim, Lukas and 
Nilles~
\cite{Chun:1992zk,Chun:1995hc} found that in models with local SUSY, i.e. supergravity, the axino mass can have a 
wider range and can be as large as the gravitino mass, $m_{3/2}$. 
Our results agree with those statements. 
Indeed a low SUSY breaking scale, for which we find a light axino, is typical of models with global SUSY,
while a higher scale, $M_{\rm SB} \geq f_a$, for which we find a heavier axino, is representative of supergravity. 
In the latter case we can identify the scale of our soft terms with the gravitino mass, 
$M_{\rm SUSY} \sim m_{3/2} \sim \frac{F}{M_p}$, with $M_p$ the Planck mass.

We emphasize, however, that the distinction between models
 of global SUSY breaking and supergravity is not strictly related to the scale 
$M_{\rm SB}$. Recently, for example, gauge mediation models with a high scale, $M_{\rm SB}>f_a$,
 have been considered (see {\it e.g.}~\cite{Zheng:2013lga}). Our statements on the axino mass only refer to the 
relative 
 size of the scales $M_{\rm SB}$ and $f_a$ and make no explicit reference to the SUSY breaking mechanism.
 
 Our results apply as long as the scales $f_a$ and $M_{\rm SB}$ are
well separated. The case $f_a \simeq M_{SB}$ would require a
special treatment because it is no longer valid to consider the SUSY
breaking effects to appear/disappear instantaneously. However this is
not possible in a momentum independent renormalization scheme like
$\overline{\text{DR}}$ which treats thresholds as step functions~\cite{Binger:2003by}, 
and is beyond the scope of this discussion here.

% 
% \begin{figure}[hbt]
% \includegraphics[width=\linewidth]{m0_A0_Q.pdf}
% \caption{Logarithm of the scale in the $(m_0,A_0)$ at which the axion soft masses become negative to trigger 
% radiative PQ breaking during the RGE running. The dashed line is for $\lambda=0.2$, the full line for $\lambda =1$.}
% \label{fig:lA0}
% \end{figure}
% We see that in this very constrained already for  $\lambda = 0.2 $ very large ratios 
% $|A_0/m_0|$ are necessary to trigger PQ breaking. However, such CMSSM scenarios are ruled out by charge 
% and color breaking minima \cite{Nilles:1982dy, Drees:1985ie,Casas:1995pd,Camargo-Molina:2013sta}. 
% Thus, if this minimal setup is realized $\lambda \geqslant 1$ is highly
% preferred, but still large valus of $|A_0|$ are needed. This would then explain automatically the 
% large mixing in the stop sector which is needed to explain the measured Higgs mass.
% The need for such large values of $A_0$ and $\lambda$ can be softened if there is a hierarchy 
% $m^2_{\chi}> m_{a}^2, m_{\bar a}^2$ in the soft masses or if the soft masses in general are significantly
% smaller than  those of the sfermions. 

\section{Axion/axino couplings to gauge fields}

The most important feature of an axion model is that it must solve the strong CP problem. This is 
achieved thanks to the fact that the $U(1)_{\rm PQ}$ is anomalous, which generates the 
following coupling of the axion to the gluons
\begin{equation} \label{eq:Lagg}
\mathcal{L}_{agg} = \frac{\alpha_s}{8 \pi} \frac{a_{\rm phys}}{f_a} G^a_{\mu\nu} \tilde G^{a \mu\nu} \, .
\end{equation}
Here $\alpha_s = \frac{g_s^2}{4 \pi}$, with $g_s$ the strong coupling constant, $a_{\rm phys}$ is the axion field, 
{\it i.e.} the massless eigenstate in the scalar CP-odd sector,
$G^a_{\mu\nu}$ the gluon field strength, and $\tilde G^{a\mu\nu} \equiv \epsilon^{\mu\nu\rho\sigma} G^a_{\rho\sigma}$. This anomalous coupling in our model is exactly the same as in the original 
non-SUSY version of the DFSZ model~\cite{Dine:1981rt}, and generates a small axion mass, $m_a$, such that $m_a f_a \simeq m_\pi f_\pi$, where $m_\pi$ and $f_\pi$ are the pion mass
and decay constant.

The SUSY model defined in the previous section departs from its non-SUSY 
counterpart~\cite{Dine:1981rt} for what concerns the coupling of axions to photons, which we
parametrize as~\cite{Beringer:1900zz}
\begin{equation} \label{eq:Lagammagamma}
\mathcal{L}_{a\gamma \gamma} = \frac{G_{a\gamma \gamma}}{4} a_{\rm phys} F_{\mu\nu} \tilde F^{\mu \nu} \, ,
\end{equation}
where
\begin{equation}
G_{a\gamma \gamma} = \frac{\alpha}{2\pi f_a} \left( \frac{E}{N} - \frac{2}{3} \frac{4 + z}{1+z} \right) \, ,
\end{equation}
with $\alpha$ the fine structure constant, $E$ and $N$ the electromagnetic and color anomalies of the 
$U(1)_{\rm PQ}$ current, $z \equiv m_u / m_d$. In the original DFSZ model~\cite{Dine:1981rt} one has
$E = \frac{4}{3} N_g (X_{H_u} + X_{H_d})$ and $N = \frac{1}{2} N_g (X_{H_u} + X_{H_d})$, which results
in $E/N = 8/3$. Here $X_{H_u}$ and $X_{H_d}$ are the PQ charges of the corresponding Higgs doublets, 
$N_g$ is the number of quark and lepton generations.
In our SUSY model there is an extra contribution to the electromagnetic anomaly which comes from
the electrically charged higgsinos. Including this contribution we find
\begin{eqnarray}
E &=& \left( \frac{4}{3} N_g + 1 \right) (X_{H_u} + X_{H_d})  \\
\frac{E}{N} & = & \frac{2}{N_g}  \left( \frac{4}{3} N_g + 1 \right) = \frac{10}{3} \, ,
\end{eqnarray}
where in the last equality we have set  $N_g = 3$. Thus the coupling to photons, which
is crucial to many experimental axion searches, is slightly modified compared to the original 
model~\cite{Dine:1981rt}.

The axino couplings to gauge fields can also be relevant, in particular to study the thermal production
of the axino in cosmology (see {\it e.g.} Refs.~\cite{Bae:2011iw, Choi:2011yf, Choi:2013lwa}). 
The form of the operators for the interactions axino - gaugino - gauge boson can be obtained via 
supersymmetrization of eq.~\eqref{eq:Lagg} and eq.~\eqref{eq:Lagammagamma}. However the coefficients
of these operators will be different as the physical axino is a slightly different linear combination of the 
fields $\hat A, \hat{\bar A}, \hat \chi, \hat H_u, \hat H_d$, compared to the axion. Such coefficients
can be calculated numerically in our model, but this is beyond the scope of the current work.

\section{Conclusion}

We have investigated what are the minimal ingredients needed to define a consistent minimal supersymmetric DFSZ axion model.
We have pointed out that the simplest model, which was first proposed in Ref.~\cite{Rajagopal:1990yx},
 is inconsistent as it suffers from a tachyonic saxion. The issue is solved by extending the 
superpotential to stabilize the PQ scale. 
We have then considered two cases: one where the SUSY breaking scale is much lower than the PQ breaking scale, 
the other where the two scales are comparable. In both cases the axion remains massless, as it should, and the 
saxion gets a mass of order $M_{\rm SUSY}$ (or $m_{3/2}$), roughly in the TeV range. 
The axino mass is dramatically different depending on the scenario. In the first ($M_{\rm SB} \ll f_a$) it is very light, 
below the keV scale, while in the second ($M_{\rm SB} \geq f_a$) it can be as large as the saxion mass. These 
results are in agreement with previous statements 
in the literature.
Furthermore, in the second case, the mixing between the new states and the MSSM Higgs sector doesn't affect the 
mass of the light Higgs but can change the prediction for the heavy Higgs mass. 

We have also discussed the couplings of the axion to gluons and photons. For the latter we found that
the presence of charged higgsinos in the SUSY model slightly modifies the strength of the coupling, which
could have implications for some experimental axion searches.

% The full model defined by the superpotential terms in Eqs.~\eqref{eq:W} and \eqref{eq:WPQ} can be defined as the 
% minimal 
% supersymmetric DFSZ axion model. It is possible at this point to make some simplifying assumptions in a CMSSM 
% fashion, 
% and study phenomenology and constraints with a relatively small parameter space. This will be the subject of future 
% work.

\section*{Acknowledgements}
We thank Branislav Poletanovi\'c for stimulating this work, Hans-Peter Nilles for clarifying discussions, and Michael 
Dine for reading a preliminary version 
of this manuscript. HD and LU 
acknowledge the DFG SFB TR 33 `The Dark Universe' for support throughout this work.

\begin{appendix}
\input{appendix} 
\end{appendix}

\bibliography{Axino}

\end{document}

%% file: appendix.tex
\section{RGEs}
\label{sec:rges}
In this appendix we give the one-loop RGEs for the model of Section~\ref{sec:consistent}. For each
parameter $X$ they are defined by the following equation:
\begin{equation}
\frac{d}{d t} X = \frac{1}{16\pi^2} \beta_X^{(1)} \, ,
\end{equation}
with $t = \log Q$, where $Q$ is the renormalization scale. For parameters already present in the MSSM we show only the difference to the corresponding RGE in the MSSM
\begin{equation}
\Delta \beta^{(1)}_X = \beta^{(1)}_X - \beta_X^{(1),MSSM} \, .
\end{equation}
The RGEs have been calculated using the generic expressions of 
Ref.~\cite{Martin:1993zk} which are implemented in {\tt SARAH} \cite{Staub:2010jh}.

\subsection*{Trilinear Superpotential Parameters}
\begin{align} 
\beta_{c_1}^{(1)} & =  
-3 c_1 g_{2}^{2}  + 3 c_1 \mbox{Tr}\Big({Y_d  Y_{d}^{\dagger}}\Big)  + 3 c_1 \mbox{Tr}\Big({Y_u  Y_{u}^{\dagger}}\Big) \nonumber \\ &  + 4 c_{1}^{2} c_1^*  + c_1 |\lambda|^2  + c_1 \mbox{Tr}\Big({Y_e  Y_{e}^{\dagger}}\Big)  -\frac{3}{5} c_1 g_{1}^{2} \\ 
\beta_{\lambda}^{(1)} & =  \lambda \Big(2 |c_1|^2  + 3 |\lambda|^2 \Big)\\ 
\Delta \beta_{Y_d}^{(1)} & =  Y_d |c_1|^2\\ 
\Delta \beta_{Y_e}^{(1)} & =  Y_e |c_1|^2 \\ 
\Delta \beta_{Y_u}^{(1)} & =   Y_u  |c_1|^2
\end{align}

\subsection*{Linear Superpotential Parameters}
\begin{align} 
\beta_{f_a^2}^{(1)} & =  f_a^2 |\lambda|^2 
\end{align}

\subsection*{Trilinear Soft-Breaking Parameters}
\begin{align} 
\beta_{T_{c1}}^{(1)} & =  
+T_{c1} \Big(12 |c_1|^2  -3 g_{2}^{2}  + 3 \mbox{Tr}\Big({Y_d  Y_{d}^{\dagger}}\Big)  + 3 \mbox{Tr}\Big({Y_u  Y_{u}^{\dagger}}\Big)  \nonumber \\ &-\frac{3}{5} g_{1}^{2}  + |\lambda|^2 + \mbox{Tr}\Big({Y_e  Y_{e}^{\dagger}}\Big)\Big)\nonumber \\ 
 &+\frac{2}{5} c_1 \Big(15 g_{2}^{2} M_2  + 15 \mbox{Tr}\Big({Y_{d}^{\dagger}  T_d}\Big)  + 15 \mbox{Tr}\Big({Y_{u}^{\dagger}  T_u}\Big)  +\nonumber \\ & 3 g_{1}^{2} M_1  + 5 \lambda^* T_{\lambda}  + 5 \mbox{Tr}\Big({Y_{e}^{\dagger}  T_e}\Big) \Big)\\ 
\beta_{T_{\lambda}}^{(1)} & =  
2 c_1^* \Big(2 \lambda T_{c1}  + c_1 T_{\lambda} \Big) + 9 |\lambda|^2 T_{\lambda} \\ 
\Delta \beta_{T_d}^{(1)} & =  |c_1|^2 T_d + 2 Y_d  c_1^* T_{c1} \\ 
\Delta \beta_{T_e}^{(1)} & =  |c_1|^2 T_e + 2 Y_e  c_1^* T_{c1} \\ 
\Delta \beta_{T_u}^{(1)} & =  |c_1|^2 T_u + 2 Y_u  c_1^* T_{c1}
\end{align}

\subsection*{Linear Soft-Breaking Parameters}
\begin{align} 
\beta_{L_V}^{(1)} & =  
\lambda^* \Big(\frac{1}{2} f_a^2 T_{\lambda}  + \lambda L_V \Big)
\end{align}

\subsection*{Soft-Breaking Scalar Masses}

\begin{align} 
\beta_{m_{ a}^2}^{(1)} & =  
2 \Big(2 \Big(m_{ a}^2 + m_{H_d}^2 + m_{H_u}^2\Big)|c_1|^2  + 2 |T_{c1}|^2 \nonumber \\
& + \Big(m_{ a}^2 + m_{\bar a}^2 + m_{\chi}^2)|\lambda|^2  + |T_{\lambda}|^2\Big)\\ 
\beta_{m_{\bar a}^2}^{(1)} & =  
2 \Big(\Big(m_{ a}^2 + m_{\bar a}^2 + m_{\chi}^2)|\lambda|^2  + |T_{\lambda}|^2\Big)\\ 
\beta_{m_{\chi}^2}^{(1)} & =  
2 \Big(\Big(m_{ a}^2 + m_{\bar a}^2 + m_{\chi}^2)|\lambda|^2  + |T_{\lambda}|^2\Big) \\
\Delta \beta_{m_{H_d}^2}^{(1)} & = +2 m_{ a}^2 |c_1|^2 +2 |T_{c1}|^2 \\ 
\Delta \beta_{m_{H_u}^2}^{(1)} & = +2  m_{ a}^2 |c_1|^2  +2 |T_{c1}|^2  
\end{align}

\subsection*{Vacuum expectation values}
\begin{align} 
\beta_{v_x}^{(1)} & =  - v_x \Big(2 |c_1|^2  + |\lambda|^2\Big)\\ 
\beta_{v_{\bar{x}}}^{(1)} & =  
- v_{\bar{x}} |\lambda|^2 \\ 
\beta_{v_{\chi}}^{(1)} & =  - v_{\chi} |\lambda|^2 \\
\Delta \beta_{v_d}^{(1)} & =  - v_d   |c_1|^2  \\ 
\Delta \beta_{v_u}^{(1)} & =  -v_u |c_1|^2 
\end{align} 